%% file: main.tex
\documentclass[%
 reprint,
 amsmath,amssymb,
 aps,
prb,
nofootinbib
]{revtex4-2}

\input{preamble.tex}

\begin{document}

\title{Orbital magnetization in two-dimensional materials from high-throughput computational screening} 

\author{Martin Ovesen}
 \email{martov@dtu.dk}
\author{Thomas Olsen}
 \email{tolsen@fysik.dtu.dk}
\affiliation{Computational Atomic-scale Materials Design, Department of Physics, Technical University of Denmark, DK-2800 Kongens Lyngby, Denmark}

\date{\today}

\begin{abstract}
We calculate the orbital magnetization of 822 two-dimensional magnetic materials from the Computational 2D Materials Database (C2DB). 
For compounds containing 5$d$ elements we find orbital moments of the order of 0.3--0.5 $\muB$, which points to the necessity of including these in any type of magnetic modeling and comparison with experiments. It is also shown that the alignment of orbital moments with respect to the spin largely follows the predictions from Hund's rule and that deviations may be explained by the $d$-band splitting originating from the crystal field - for example in the important case of CrI$_3$. Finally, we show that for certain insulators, Hubbard corrections may lead to large and fully unquenched orbital moments that are pinned to the lattice rather than the spin and that these moments can lead to enormous magnetic anisotropies. Such unquenched ground states are only found from density functional theory calculations that include both Hubbard corrections and self-consistent spin--orbit coupling and largely invalidates the use of the magnetic force theorem for calculating magnetic anisotropies.
\end{abstract}

\maketitle

\section{Introduction}
\input{Sections/Introduction.tex}

\section{Theory}\label{sec:theory}
\input{Sections/Theory.tex}

\section{Computational details}\label{sec:computational}
\input{Sections/CompDetails.tex}

\section{Results}\label{sec:results}
\input{Sections/ResultsDB.tex}
\input{Sections/ResultsDFT+U.tex}

\section{Conclusion}\label{sec:conclusion}
We have presented high-throughput calculations of orbital magnetization of 822 stable magnetic monolayers from the C2DB. The results have been added to the C2DB and may be browsed freely online or are available for download upon request. As expected, the largest orbital moments are found in heavy elements such as Pt, Au and Os, where they may reach 0.3--0.5 $\muB$. It was shown that the atomic picture of alignment/anti-alignment based of $d$-band filling is satisfied to some extent, but also show significant deviations for the 3$d$ metals with $d$-bands less than half-filled. In particular, the important case of Cr atoms are often predicted to exhibit aligned spin and orbital moments although Hund's rule would predict anti-alignment. Such deviations can be understood from crystal field splitting, which splits the $d$-band into multiplets and it is often the occupation of these that determines the sign of orbital magnetization. We also showed that no apparent correlation exists between orbital magnetization and magnetic anisotropy despite the fact that both are governed by SOC effects. Finally, it was shown that for certain monolayers, Hubbard corrections may drive the splitting of partially occupied $t_{2g}$ bands, leading to fully unquenched orbital moments - even in systems where strong SOC is not expected. This mechanism is not captured by the PBE+NSCF SOC applied in the workflow and our reported orbital moments in such systems are not expected to be reliable. On the other hand, for such cases the basic PBE electronic structure without Hubbard corrections (as reported in the C2DB) is not expected to be reliable either.

The orbital magnetization plays an important role for several magnetic properties. For example, the longitudinal magnetoelectric effect is typically completely dominated by orbital effects \cite{malashevich2012full} and the orbital Hall effect has been proposed to comprise a key ingredient in understanding the microscopic origin in the spin Hall effect \cite{Kontani2009}. Finally, the fully unquenched orbital moments found in VI$_3$ and FePS$_3$ are pinned to the lattice rather than the spin and thus lead to exceedingly large magnetic anisotropies. Since magnetic anisotropy plays a pivotal role for magnetic order in 2D \cite{mermin1966absence}, the design of materials with fully unquenched orbital moments could lead to the discovery of new 2D magnets with large critical temperatures.

\bibliography{bibtex.bib}

\end{document}

%% file: preamble.tex
\usepackage{graphicx}
\usepackage{dcolumn}
\usepackage{bm}
\usepackage{amssymb} 
\usepackage{multirow}
\usepackage{amsmath}
\usepackage{mathtools}
\usepackage[dvipsnames,svgnames,x11names,hyperref]{xcolor}
\usepackage[pdfusetitle]{hyperref}
\usepackage{etoolbox}
\apptocmd{\sloppy}{\hbadness 3000\relax}{}{} 

\hypersetup{
  colorlinks,
  citecolor=blue,
  linkcolor=blue,
  urlcolor=black}

\makeatletter
\newcommand{\fmarki}{*}
\newcommand{\fmarkii}{\ensuremath{\dagger}}
                
\def\@fnsymbol#1{{\ifcase#1\or \fmarki\or \fmarkii \else\@ctrerr\fi}}
\makeatother
\renewcommand{\fmarki}{\ensuremath{\dagger}}
\renewcommand{\fmarkii}{\ensuremath{\dagger\dagger}}

\def\Vhrulefill{\leavevmode\leaders\hrule height 0.7ex depth \dimexpr0.4pt-0.7ex\hfill\kern0pt}

\newcommand{\bra}[1]{\left<\vphantom{\sum}#1\right|}
\newcommand{\ket}[1]{\left|\vphantom{\sum}#1\right>}
\newcommand{\braket}[2]{\left<\vphantom{\sum}#1\middle|#2\right>}
\newcommand{\brK}[1]{\left[#1\right]}
\newcommand{\brB}[1]{\hspace{-1.5 pt}\left(#1\right)}
\newcommand{\brN}[1]{\left|#1\right|}

\newcommand{\muB}{\mu_\mathrm{B}}

\newcommand{\DArrow}{\scalebox{0.6}{$\downarrow$}}
\newcommand{\UArrow}{\scalebox{0.6}{$\uparrow$}}

\renewcommand{\vec}[1]{\mathbf{#1}}

%% file: Sections/Introduction.tex
The microscopic origin of magnetism is typically understood in terms of Pauli exclusion and electronic Coulomb repulsion \cite{yosida1996theory}. For insulators containing transition metals the magnetic moment per unit cell can often be predicted from the oxidation state of the magnetic atoms along with Hund's rule and the resulting magnetic order may be ferromagnetic, collinear antiferromagnetic or helical depending on the orbital hybridization. In fact, in the absence of spin-orbit coupling (SOC) collinear insulators are guaranteed to host an integer number of Bohr magnetons per unit cell originating from the electronic spin.  For example, the Cr$^{3+}$ state of Cr atoms in either Cr$_2$O$_3$ of CrI$_3$ leads to three unpaired occupied $d$-states and a magnetization of three Bohr magnetons per Cr atom. Such predictions are readily verified by, for example, density functional theory (DFT) calculations, but for the majority of magnetic insulators it is not really necessary to perform any calculations if one simply wants to estimate the magnetization. The situation is somewhat more complicated for metals where the order is best understood from the Stoner model \cite{yosida1996theory}, which does not yield a qualitative estimate of the magnetization in real materials. However, in that case the spin magnetization is straightforward to calculate with DFT, which typically provides good agreement with experiments \cite{bihlmayer2018density}).

When SOC is taken into account, the spin along the magnetization direction is not a good quantum number and the magnetization will deviate from integer values in insulators. In addition, orbital magnetization may contribute to the total magnetization and could potentially yield large contributions that need to be taken into account for accurate magnetic modeling. In general the calculation of orbital magnetization in interacting systems requires either many-body theory \cite{aryasetiawan2016green} or current density functional theory \cite{vignale1995current} and there seems to have been rather limited progress in {\it ab initio} approaches towards this. Nevertheless, it is possible to calculate the orbital magnetization of non-interacting systems and in the framework of DFT it is often assumed that the orbital magnetization of the Kohn-Sham system provides a reasonable estimate for the true orbital magnetization. However, even for non-interacting systems the theoretical treatment of orbital magnetization is non-trivial for periodic systems due to subtleties involved with the angular momentum operator \cite{thonhauser2005orbital,ceresoli2006orbital,thonhauser2011theory}. This has largely been solved with the advent of the "modern theory" of orbital magnetization where a closed expression has been derived using either a semiclassical approach \cite{xiao2005berry}, Wannier functions \cite{thonhauser2005orbital,ceresoli2006orbital}, or linear response methods \cite{aryasetiawan2019modern}. Alternatively, the orbital magnetization can be estimated by assuming localized orbital moments centered on the atoms and simply resolving the Bloch states in terms of atom-centered contributions \cite{todorova2001current}. While the former approach is expected to be more accurate, the latter makes it possible to resolve the orbital magnetization into contributions from individual atoms. Both methods have been shown to yield reasonable agreement with experiments for simple ferromagnetic metals \cite{ceresoli2010first}, while only the modern theory is able to predict the large orbital magnetization induced by the non-coplanar ground state in a monolayer of Mn \cite{hanke2016role}. 

The orbital magnetization of bulk Fe, Ni, and Co has been studied in detail with various computational techniques and constitutes 5--10 \% of the total magnetization \cite{ceresoli2010first}, which is in good agreement with experiments based on the Einstein--de Hass effect \cite{meyer1961experimental}. While 5--10 \% may in itself be regarded as a significant contribution, the effect of orbital magnetization could potentially be much larger in materials with stronger SOC. Moreover, certain insulators (without strong SOC) have been shown to host orbital magnetic moments of up to one Bohr magneton per magnetic atom which is driven by Hund´s rule splitting of the $t_{2g}$ manifold \cite{hovancik2023large}. Nevertheless, in most computational studies of magnetic materials the orbital magnetization is not reported - most likely due to (sometimes) unjustified assumptions of small contributions as well as the subtleties involved in the extraction of the orbital magnetization from DFT. The inclusion of orbital magnetization can nonetheless be of crucial importance when interpreting measured magnetic susceptibilities and plays a vital role for magnetoelectric effects in certain materials \cite{malashevich2012full,mangeri2024linear}.

In recent years there has been a rapidly growing interest in two-dimensional (2D) magnetic materials \cite{gibertini2019magnetic}. Starting with the experimental discovery of ferromagnetism in CrI$_3$ \cite{huang2017layer}, there is now wide variety of 2D materials where magnetic order has been demonstrated in the monolayer limit at low temperature. Examples includes itinerant ferromagnetism in Fe$_3$GeTe$_2$ \cite{fei2018two}, 
collinear antiferromagnetism in FePS$_3$ \cite{lee2016ising}, itinerant antiferromagnetism in CrTe$_2$ \cite{xian2022spin} and spiral order leading to multiferroicity in NiI$_2$ \cite{song2022evidence}. In addition, a large number of computational studies have reported predictions of magnetic order in 2D materials from high throughput screening of materials databases \cite{mounet2018two,torelli2019high,torelli2020high,shen2022high}. Except for a few specific 2D materials \cite{hanke2016role,Yang2020,bikaljevic2021noncollinear,sandratskii2021interplay,hovancik2023large,lee2023giant}, the effect of orbital magnetization have largely been neglected in these studies and a systematic screening of orbital magnetization in 2D has been lacking.

In the present work we report a high throughput computational screening of orbital magnetization in 2D materials using DFT with non-self-consistent SOC calculations. We calculate the orbital magnetization of 822 magnetic materials from the Computational 2D Materials Database (C2DB) and analyze the results in terms of Hund's rule and magnetic anisotropy. As expected, we find the largest orbital moments (up to 0.5 Bohr magnetons) for heavy atoms, but do not find any correlation with magnetic anisotropy despite the fact that both effects are driven by SOC. For the 3$d$ transition metals we find that the alignment of orbital moments with respect to the spin largely follow the expectation from Hund's rule for elements where the $d$-shell is more than half full, but less so for elements with sparsely populated $d$-shells. In particular, for the important case of magnetic compounds containing Cr atoms, we find both aligned and anti-aligned moments, which largely contradicts Hund's rule predicting anti-aligned orbital moments. All results have been added to the C2DB \cite{haastrup2018computational} and may be browsed online or downloaded freely upon request. Finally, we show that for certain insulators it is crucial to perform self-consistent SOC calculations to obtain the correct orbital magnetization. In such cases, the orbital magnetization can become on the order of one Bohr magneton and we comment on why and when such failure of non-self-consistent SOC is expected.

The paper is organized as follows. In Sec.~\ref{sec:theory} we outline the basic theory of orbital magnetization as well as the computational approach applied in this work. In Sec.~\ref{sec:computational} we present the computational details for the DFT calculations as well as the workflow applied for the high throughput computations. In Sec.~\ref{sec:results} the results are presented and we analyze the extend to which Hund's rule for the sign of orbital moments is satisfied and show that correlations between magnetic anisotropy and orbital magnetization is absent. We then discuss specific cases where the non-self-consistent approach fails and exemplify the origin of such failures for VI$_3$ and FePS$_3$. Sec.~\ref{sec:conclusion} provides a discussion and outlook.

%% file: Sections/Theory.tex
We start by briefly summarizing the approach for calculating orbital magnetic moments in non-interacting systems. For a finite system, the total orbital magnetic moment may be related to the expectation value of the orbital angular momentum operator $\hat{\vec{L}}=-i\hbar\hat{\vec{r}}\times\hat{\boldsymbol{\nabla}}$. Given a set of two-component single-particle eigenstates $\Psi_n=\brK{\psi_{\UArrow n},\psi_{\DArrow n}}^T$ it may thus may thus be calculated as 
\begin{equation}
    \vec{m}_\mathrm{orb}=-\frac{\muB}{\hbar}\sum_{\sigma n}f_{n}\bra{\psi_{\sigma n}}\hat{\vec{L}}\ket{\psi_{\sigma n}}\hspace{-2 pt},
    \label{eq:OrbMagFinite}
\end{equation}
 where $f_n$ is the thermal occupation factor corresponding to the eigenstate $\Psi_n$.

For an extended system with a unit cell volume $V$, it may seem sensible to calculate the orbital magnetization (orbital moment per volume) through a similar equation
\begin{equation}
    \vec{M}_\mathrm{orb}=-\frac{\muB}{\hbar V}\sum_{\vec{k}\sigma n}f_{\vec{k}n}\bra{\psi_{\vec{k}\sigma n}}\hat{\vec{L}}\ket{\psi_{\vec{k}\sigma n}}\hspace{-2 pt},
    \label{eq:OrbMagPeriodic}
\end{equation}
where $f_{\vec{k}n}$ is now the thermal occupation factor multiplied by the weight of the $\vec{k}$-point. However, the action of the position operator that appears in $\hat{\vec{L}}$ is not well-defined on the periodic Bloch functions and the expression is ill-defined as it stands. This problem was resolved in the formulation of the modern theory of orbital magnetization \cite{xiao2005berry,thonhauser2005orbital,ceresoli2006orbital,shi2007quantum} where a simple Berry-phase type of expression was derived for the orbital magnetization using Wannier functions. Alternatively, one may apply an atom-centered approximation (ACA) where the orbital magnetisation is written as a sum of local atomic orbital magnetic moments: 
\begin{equation}
    \vec{M}_\mathrm{orb}=\frac{1}{V}\sum_a\vec{m}_\mathrm{orb}^a,
\end{equation}
where the sum runs over atoms in the unit cell. Typically when the ACA is formulated, the matrix elements yielding the orbital magnetic moments are restricted to spherical regions $\Omega_a$ centered at each atomic position $\vec{R}_a$ with some cutoff radius such that the atomic orbital magnetic moments can be calculated through
\begin{equation}
    \vec{m}_\mathrm{orb}^a=i\muB\sum_{\vec{k}\sigma n}f_{\vec{k}n}\bra{\psi_{\vec{k}\sigma n}}\hat{\vec{r}}_a\times\hat{\boldsymbol{\nabla}}\ket{\psi_{\vec{k}\sigma n}}_\mathrm{\hspace{-2 pt}\Omega_a}\hspace{-3 pt},
\label{eq:morb_aca}
\end{equation}
where $\hat{\vec{r}}_a=\hat{\vec{r}}-\vec{R}_a$ is the position operator with the origin shifted to the center of atom $a$. The underlying assumption of this approach is that orbital moments are dominated by localized atomic orbitals (typically $d$-orbitals), and that interstitial contributions can be neglected. For the simple ferromagnets, Fe, Ni and Co, the ACA yields reasonable agreement with the modern theory, but has been shown to deviate significantly in the case of a monolayer Mn, which exhibits non-coplanar magnetic order \cite{hanke2016role}. We expect, however, that the ACA may provide reasonable estimates for the orbital magnetization in most collinear systems. In addition, the ACA has the advantage that orbital magnetization is easily decomposed into contributions from individual atoms and orbitals, which becomes crucial when analyzing orbital magnetization effects in antiferromagnets. Finally, we emphasize that both the modern theory and ACA yields the non-interacting orbital magnetization (of the Kohn-Sham system) and thus represent rough estimates of the true orbital magnetization.

In the present work we have applied the ACA approach within the projector augmented--wave (PAW) framework \cite{blochl1994projector} where the all-electron wave functions are expressed as
\begin{equation}
\begin{split}
    \psi_{\vec{k}\sigma n}\brB{\vec{r}}=\widetilde{\psi}_{\vec{k}\sigma n}\brB{\vec{r}}+\sum_a\sum_i&\braket{\widetilde{p}^a_i}{\widetilde{\psi}_{\vec{k}\sigma n}} \\
    &\hspace{4 pt} \times\brK{\phi^a_i\brB{\vec{r}_a}-\widetilde{\phi}^a_i\brB{\vec{r}_a}}\hspace{-2 pt}.
\end{split}
\end{equation}
Here $\widetilde{\psi}_{\vec{k}\sigma n}$ is the smooth pseudo-wave function and $\widetilde{p}^a_i$ denote the projector functions which project $\widetilde{\psi}_{\vec{k}\sigma n}$ onto the basis of all-electron partial waves $\phi^a_i$ and pseudo-partial waves $\widetilde{\phi}^a_i$. Inside the PAW spheres the $\widetilde{\psi}_{\vec{k}\sigma n}$ cancels the contributions from the pseudo-partial waves and in the vicinity of atom $a$ the all-electron wave function may be written as
\begin{equation}
    \psi_{\vec{k}\sigma n}\brB{\vec{r}}=\sum_i\braket{\widetilde{p}^a_i}{\widetilde{\psi}_{\vec{k}\sigma n}}\phi^a_i\brB{\vec{r}_a}\quad\mathrm{if}\quad\vec{r}\in\Omega_a,
\label{eq:paw_close}
\end{equation}
where $\Omega_a$ denotes the spherical augmentation region surrounding atom $a$.

This framework lends itself especially well to calculating local orbital magnetic moments within the ACA. By considering only contributions to the moments close to the atom, equation~\eqref{eq:paw_close} can be inserted into equation~\eqref{eq:morb_aca} yielding
\begin{equation}
\begin{split}
    \vec{m}_\mathrm{orb}^a=i\muB\sum_{i_1i_2}&
    \bra{\phi^a_{i_1}}\hat{\vec{r}}_a\times\hat{\boldsymbol{\nabla}}\ket{\phi^a_{i_2}}_{\Omega_a} \\
    &\hspace{6 pt}\times\sum_{\vec{k}\sigma n}f_{\vec{k}n}\braket{\widetilde{p}^a_{i_1}}{\widetilde{\psi}_{\vec{k}\sigma n}}^*\hspace{-3 pt}\braket{\widetilde{p}^a_{i_2}}{\widetilde{\psi}_{\vec{k}\sigma n}}\hspace{-2 pt}.
\end{split}
\end{equation}
By defining the elements of the atomic density matrix as
\begin{equation}
    D^a_{i_1i_2}=\sum_{\vec{k}\sigma n}f_{\vec{k}n}
    \braket{\widetilde{p}^a_{i_1}}{\widetilde{\psi}_{\vec{k}\sigma n}}^*\hspace{-3 pt}
    \braket{\widetilde{p}^a_{i_2}}{\widetilde{\psi}_{\vec{k}\sigma n}}
\end{equation}
and the elements of the angular momentum operator w.r.t. the partial waves as
\begin{equation}
    \vec{L}^a_{i_1i_2}=-i\bra{\phi^a_{i_1}}\hat{\vec{r}}_a\times\hat{\boldsymbol{\nabla}}\ket{\phi^a_{i_2}}_{\Omega_a}\hspace{-3 pt},
\end{equation}
the atomic orbital moments can be written as
\begin{equation}
    \vec{m}_\mathrm{orb}^a=-\muB\sum_{i_1i_2}
    D^a_{i_1i_2}\vec{L}^a_{i_1i_2}.
\label{eq:orbmag_final}
\end{equation}
When calculating the matrix elements $\vec{L}^a_{i_1i_2}$, it is beneficial to split the partial waves into their angular part, described by the \textit{real} spherical harmonics, and their radial part
\begin{equation}
    \phi^a_i\brB{\vec{r}_a}=R^a_{nl}\brB{r_a}Y^a_{lm}\brB{\theta_a,\varphi_a},
\end{equation}
where $i$ is a composite index for the usual quantum numbers $n$, $l$ and $m$ of a hydrogen-like atom. We then get
\begin{equation}
\begin{split}
    \vec{L}^a_{i_1i_2}=-i\bra{Y^a_{l_1m_1}}\hat{\vec{r}}_a\times\hat{\boldsymbol{\nabla}}&\ket{Y^a_{l_2m_2}} \\
    &\hspace{19 pt}\times\braket{R^a_{n_1l_1}}{R^a_{n_2l_2}}_{\Omega_a}\hspace{-4 pt}.
\end{split}
\label{eq:angmom_elements}
\end{equation}
Note that both the elements $D^a_{i_1i_2}$ and $\vec{L}^a_{i_1i_2}$ separately produce Hermitian matrices and that the elements $\vec{L}^a_{i_1i_2}$ are either zero or strictly imaginary. Consequently when the sum in equation \eqref{eq:orbmag_final} is performed, the real parts of $D^a_{i_1i_2}$ all cancel out exactly and only the imaginary parts contribute.  This a manifestation of the well-known fact that orbital magnetization is quenched in the absence of SOC for coplanar systems (in which case the orbitals may be chosen as real).
\begin{table}[t]
\begin{ruledtabular}
\def\arraystretch{1.3}
\begin{tabular}{l|llll}
Crystal & \multicolumn{1}{c}{\textbf{bcc-Fe}} & \multicolumn{1}{c}{\textbf{hcp-Co}}
        & \multicolumn{1}{c}{\textbf{fcc-Co}} & \multicolumn{1}{c}{\textbf{fcc-Ni}} \\
Easy axis & \multicolumn{1}{c}{$[001]$} & \multicolumn{1}{c}{$[001]$}
          & \multicolumn{1}{c}{$[111]$} & \multicolumn{1}{c}{$[111]$} \\[2pt] \hline
PAW-ACA                                 & $0.0611$ & $0.0886$ & $0.0845$ & $0.0546$ \\
MT-ACA\cite{ceresoli2010first}          & $0.0433$ & $0.0868$ & $0.0634$ & $0.0511$ \\
Modern theory\cite{ceresoli2010first}   & $0.0658$ & $0.0957$ & $0.0756$ & $0.0519$ \\
Measurement\cite{meyer1961experimental} & $0.081$  & $0.133$  & $0.120$  & $0.053$
\end{tabular}
\caption{Calculated and measured values of the orbital magnetic moments in units of $\muB$ per atom.}
\label{tab:orbmag_benchmark}
\end{ruledtabular}
\end{table}
The expansion \eqref{eq:paw_close} is only exact inside the PAW spheres whereas the inner product $\braket{R^a_{n_1l_1}}{R^a_{n_2l_2}}$ is normalized in a region that extends beyond the PAW spheres. Nevertheless, we will take $\braket{R^a_{n_1l_1}}{R^a_{n_2l_2}}\approx\delta_{j_1j_2}$ in Eq.~\eqref{eq:angmom_elements}, which corresponds to enforcing a renormalization of the partial waves inside the PAW spheres. Alternatively this can be viewed as extending the expression  \eqref{eq:paw_close} to regions beyond the PAW spheres such that normalization of the partial waves are guaranteed.

Our methodology, the PAW atom-centered approximation (PAW-ACA), has been benchmarked against the simple ferromagnetic metals bcc-Fe, hcp-Co, fcc-Co and fcc-Ni. The calculated values for $|\vec{m}_\mathrm{orb}^a|$ are displayed in \autoref{tab:orbmag_benchmark} alongside the observed value obtained by measurement of the Einstein--de Haas effect and values obtained through the muffin-tin atom-centered approximation (MT-ACA) and the modern theory. For all compounds, the orbital magnetization is aligned with the spin magnetization. The benchmark shows that the simple methodology presented here provides good agreement with the modern theory of orbital magnetization although both approaches underestimate the measured values.

%% file: Sections/CompDetails.tex
The approach for calculating orbital magnetic moments described in Sec.~\ref{sec:theory} was applied to the Computational 2D Materials Database (C2DB) \cite{haastrup2018computational,gjerding2021recent} which constitutes a collection of 16818 two-dimensional materials obtained through various methods such as exfoliation of known layered van der Waals bulk crystals, lattice decoration protocols and crystal diffusion variational auto-encoder models \cite{lyngby2022data}. In this work, only the dynamically and thermodynamically stable magnetic materials in C2DB have been selected, which results in 822 monolayers.
\begin{figure}[b]
    \centering
    \includegraphics[width=1\linewidth]{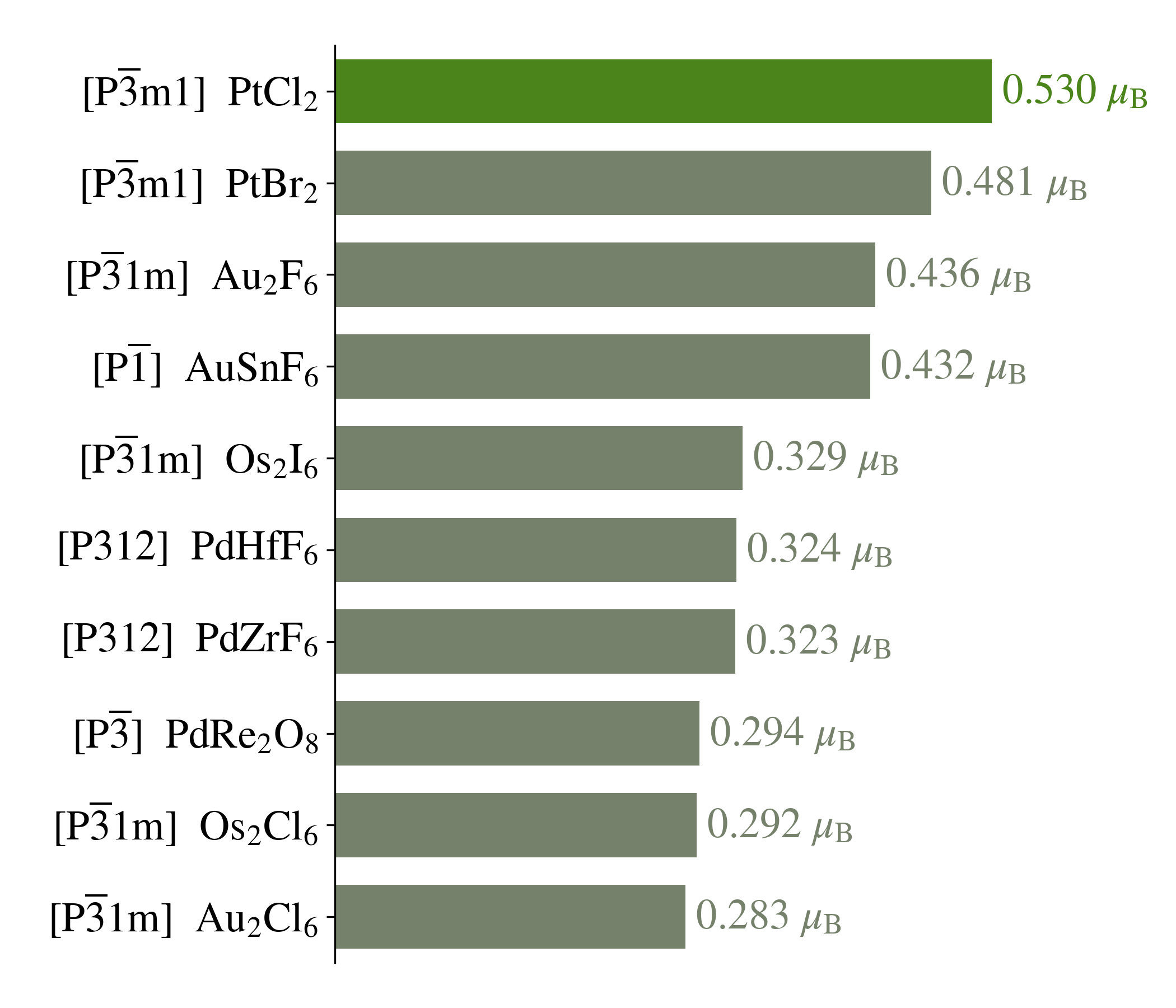}
    \vspace{-10 pt}\caption{Space groups and atoms in the primitive unit cell of crystals in C2DB with the largest $\brN{m_\mathrm{orb}^a}$.}
    \label{fig:largest_om}
\end{figure}
All calculations were carried out using density functional theory as implemented in the open-source electronic structure package GPAW \cite{mortensen2005,enkovaara2010,mortensen2024gpaw} in combination with the Atomic Simulation Environment (ASE) \cite{larsen2017atomic}. We used the Perdew-Burke-Ernzerhof (PBE) exchange-correlation functional \cite{perdew1996generalized}, a plane wave cutoff of 800 eV, and a $\vec{k}$-point density of 12 Å. The spin--orbit coupling  was included non-self-consistently (NSCF SOC) \cite{olsen2016designing} with collinear magnetization aligned with the easy-axis (also determined from non-self-consistent SOC) \cite{vannucci2020anisotropic}.
The matrix elements of the angular momentum operator were calculated in accordance with equation~\eqref{eq:angmom_elements} and the atomic orbital magnetic moment vector was calculated through equation~\eqref{eq:orbmag_final}. In order to characterize the orbital moment for a large amount of materials, the vector was converted to a simple scalar function by taking the dot product with the unit vector pointing along the magnetic easy axis. This yields a signed measure denoted as $m_\mathrm{orb}^a$ which is positive when the spin and orbital moments are aligned and negative when they are anti-aligned.

%% file: Sections/ResultsDB.tex
\subsection{High-throughput calculations}

All results have been added to the C2DB and the orbital moments for all stable magnetic materials can be accesses from the database or queried with a browser online. We start by reporting the most extreme cases - that is, the materials exhibiting the highest orbital moments. As shown in \autoref{fig:largest_om}, the materials with the largest moments mainly have heavy 5d transition metal constituents such as Pt, Au, and Os although Pd also appears. This is generally expected as orbital magnetization emerges solely from SOC in the case of coplanar magnetic ordering \cite{hanke2016role}, and the spin--orbit interaction energy generally increases with the atomic number. 
\begin{figure*}
    \centering
    \includegraphics[width=1\textwidth]{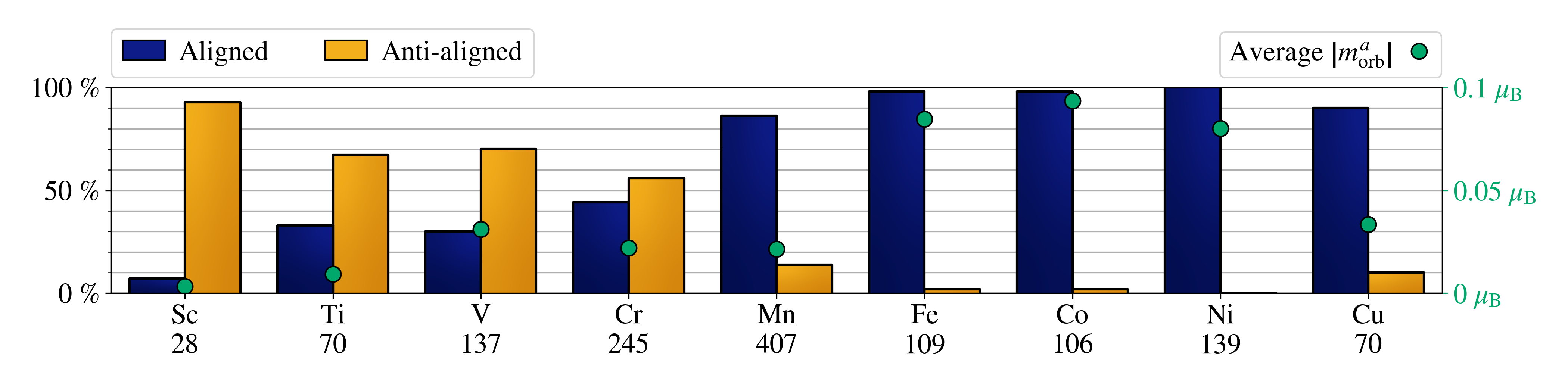}
    \caption{Distribution of aligned and anti-aligned magnetic moments for the 3$d$ elements of stable magnetic monolayers from the C2DB. The lowermost numbers indicate the occurrences of each element in the database and the green points mark the average norm of the atomic orbital moment.}
    \label{fig:HundTest}
\end{figure*}
The orbital magnetization is typically reported as being collinear with the spin magnetization\cite{meyer1961experimental,frisk2018magnetic,fujita2022x,watson2020direct,hovancik2023large,bikaljevic2021noncollinear,lee2023giant}, but the spin and orbital contributions may be aligned or anti-aligned. Hund's rules serve as a simple model that describes the electron configuration of a given atom; the third rule essentially stating that the spin and orbital magnetic moments should be anti-aligned when the subshell of an atom is less than half-filled and aligned when the subshell is more than half-filled. As a test of Hund's third rule, the orbital moments of each 3$d$ atom present in our database has been collected and sorted as either aligned or anti-aligned with the spin moment. The resulting distributions are presented in \autoref{fig:HundTest}, displaying excellent agreement for iron, cobalt and nickel especially. Additionally, the 3$d$ atoms with less than half-filled subshells also tend to prefer anti-aligned local moments in agreement with Hund's third rule; however, they do not display the same dominating proclivity towards anti-alignment as would be expected. Especially Cr which seems to only have a 54 \% fraction of anti-alignment.
\begin{figure}[tb]
    \centering
    \includegraphics[width=1\linewidth]{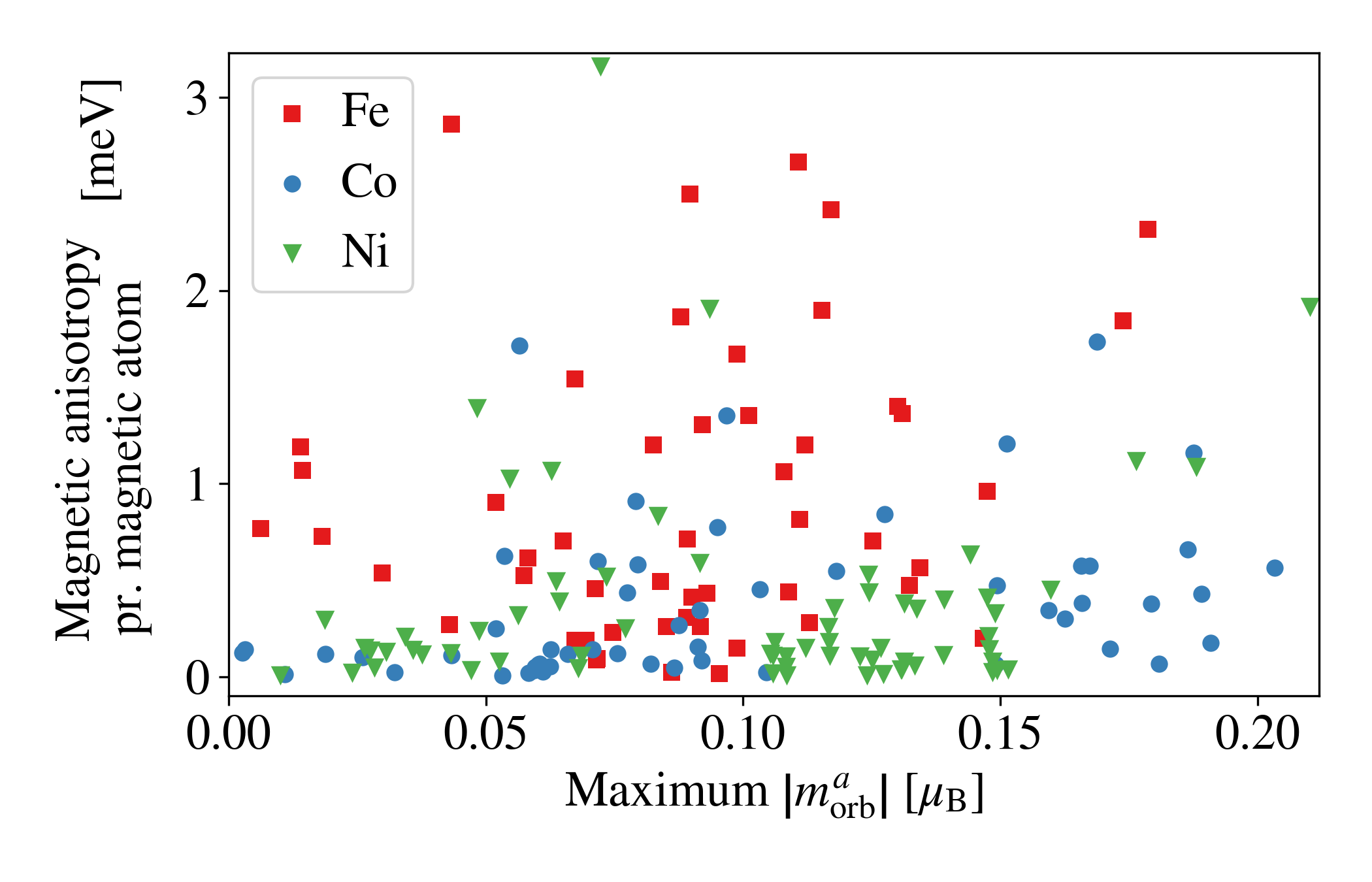}
    \caption{Magnetic anisotropy against the maximum orbital magnetic moments of magnetic crystals in C2DB. Respectively, red squares, blue circles and gray triangles denote crystals with Fe, Co and Ni as the magnetic atoms. The plot has been zoomed in on the largest mass of data points although this hides some crystals with large magnetic anisotropy. These include two Fe compounds at (0.053 $\muB$, 4.79 meV), (0.085 $\muB$, 5.26 meV), one
    Co compound at (0.117 $\muB$, 4.75 meV), and three
    Ni compounds at (0.042 $\muB$, 12.22 meV), (0.073 $\muB$, 4.92 meV) and (0.174 $\muB$, 4.67 meV).}
    \label{fig:UvsMagAni}
\end{figure}

In \autoref{fig:HundTest} we also show the average magnitude of the moments for each set of materials containing a particular 3$d$ element. Again, the orbital magnetization is expected to depend on the overall strength of SOC, which increase with atomic number. We do see a slight tendency for increasing moments as the atomic number increases across the 3$d$ series, but it is highly likely that ligand effects are equally important for SOC in this set of compounds. In addition, the spin state of the magnetic atoms plays an even more crucial role. For example, Mn typically resides in an oxidation state that yields a half-filled $d$-shell with $S=5/2$. In that case the orbital momentum would be zero for an isolated atom and the orbital magnetization in Mn compounds is small, albeit not completely vanishing due to hybridization.

Finally, the crystal field splitting will often give rise to much larger energy shifts than those induced by SOC and can lead to significant modifications of the spin/orbital moment alignment as predicted by Hund's rule. For example, the case of CrI$_3$ \cite{huang2017layer} exhibits octahedral crystal field splitting and $S=3/2$ spin states, which lead to a filled $t_{2g}$ band for the majority spin. As such, no significant orbital magnetization is expected and the small computed values are primarily due to ligand effects and hybridization.

Magnetic anisotropy is another emerging spin--orbit effect; consequently, it is reasonable to expect correlation between the magnitudes of orbital magnetisation and magnetic anisotropy. In particular, for cases where the magnetic anisotropy is dominated by single-ion anisotropy, the easy-axis can often be predicted from crystal field splitting combined with a filling of orbital angular momentum levels \cite{Yiu2017}. While it may be difficult to quantify the {\it magnitude} of anisotropy in such cases, it is expected to be rather strongly correlated with orbital magnetization. In order to test this correlation, the materials where iron, nickel or cobalt are the main magnetic atoms have been extracted from C2DB and the magnetic anisotropy per magnetic atom vs. the maximum $|m_\mathrm{orb}^a|$ in the unit cell has been plotted in \autoref{fig:UvsMagAni}. No immediate correlation can be established as our database exhibits materials with both small and large anisotropies along the whole range of orbital moment magnitudes. Both orbital moments and magnetic anisotropy are subject to ligand effects, but the present results showcases the difficulty in predicting either quantity from atomic numbers (and thus the strength of SOC) of the magnetic atoms alone.

%% file: Sections/ResultsDFT+U.tex
\subsection{Interplay between DFT\texttt{+}U and spin--orbit}

\begin{table*}
\begin{ruledtabular}
\def\arraystretch{1.3}
\begin{tabular}{lccccr}
\multicolumn{1}{c}{\multirow{2}{*}{Crystal}} & PBE & PBE+U & LSDA+U & LSDA+U & \multicolumn{1}{c}{\multirow{2}{*}{Measurement}} \\
         & NSCF SOC & NSCF SOC & NSCF SOC & SCF SOC & \\[-4pt]
 \multicolumn{6}{c}{\Vhrulefill} \\[-2pt]
 CrI$_3$    & \phantom{-}0.099 & 0.055 & \phantom{-}0.097 & \phantom{-}0.114 & -0.067 \cite{frisk2018magnetic}\footnotemark[1] \\
 CrGeTe$_3$ & \phantom{-}0.058 & 0.031 & \phantom{-}0.010 & \phantom{-}0.062 & -0.022 \cite{watson2020direct}\footnotemark[1] \\
 CrSiTe$_3$ & \phantom{-}0.016 & 0.010 & \phantom{-}0.014 & \phantom{-}0.023 & -0.033 \cite{fujita2022x}\footnotemark[2] \\
 NiBr$_2$   & \phantom{-}0.134 & 0.062 & \phantom{-}0.069 & \phantom{-}0.184 & 0.34 to 0.41 \cite{bikaljevic2021noncollinear}\footnotemark[2] \\
 VI$_3$     & -0.067           & 0.026 & -0.033 & \textbf{-0.953} & -0.7 to -0.5 \cite{hovancik2023large}\footnotemark[2] \\
 FePS$_3$   & \phantom{-}0.105 & 0.042 & \phantom{-}0.069 & \textbf{\phantom{-}0.847} & 0.98 to 1.06 \cite{lee2023giant}\footnotemark[1]
\end{tabular}
\end{ruledtabular}
\footnotetext[1]{Value obtained through comparison between experimental and simulated x-ray absorption spectra (XAS).}
\footnotetext[2]{Value obtained through magnetic sum rules of x-ray magnetic circular dichroism (XMCD) spectra.}
\caption{Calculated and measured local orbital magnetic moments for known van der Waals-layered materials. Values are in units of $\muB$ pr. magnetic (3$d$) atom. DFT+U calculations have all been performed with a plane wave energy cutoff of 800 eV, a $24\times24$ $\vec{k}$-point grid, and a Hubbard penalty $U=3$ eV on the $d$ orbitals of the 3$d$ elements. Results in bold mark a dramatic increase in the local orbital moment.}
\label{tab:vdw-compare}
\end{table*}

While \autoref{tab:orbmag_benchmark} showed that the PAW-ACA method (with NSCF SOC) yielded good agreement with measurements on simple bulk metals, there are also examples where the approach fails rather dramatically. Here we will aim to compare and discuss deviations between our calculations and measurements of the orbital magnetic moments in few-layered van der Waals--crystals. The calculated values are displayed in \autoref{tab:vdw-compare} under the PBE NSCF SOC tab. For the chromium compounds, CrI$_3$, CrSiTe$_3$, and CrGeTe$_3$, the measured orbital moments are generally small ($<0.1\ \mu_B$) and anti-aligned with the spin moment. The computed orbital moments, however, are all predicted to be aligned with the spin moment. It is not obvious though, that Hund's rule is applicable since all of these monolayers exhibit octahedral field splitting and a filled $t_{2g}$ band implying vanishing orbital magnetization. The remaining (small) unquenched moments originate from deviations from octahedral field splitting as well as SOC-mediated mixing between $e_g$ and $t_{2g}$ states, which does not yield a simple prediction for the direction of the moments. Nevertheless, the discrepancy signals a deficiency in the DFT description of these compounds at the PBE level.

Measurements of the orbital moments have also been reported for hexagonal NiBr$_2$, VI$_3$, and FePS$_3$. Experimentally the latter two exhibit very large local moments and while the computed moments from C2DB have the correct alignment, they are all significantly underestimated, especially for VI$_3$ and FePS$_3$ where the calculated local moments are an order of magnitude too small. Additionally, both of these crystals are experimentally known as semiconductors \cite{du2016weak,kong2019vi3}, but are predicted to be metallic in the C2DB.
\begin{figure}[tb]
    \centering
        \includegraphics[width=1\linewidth]{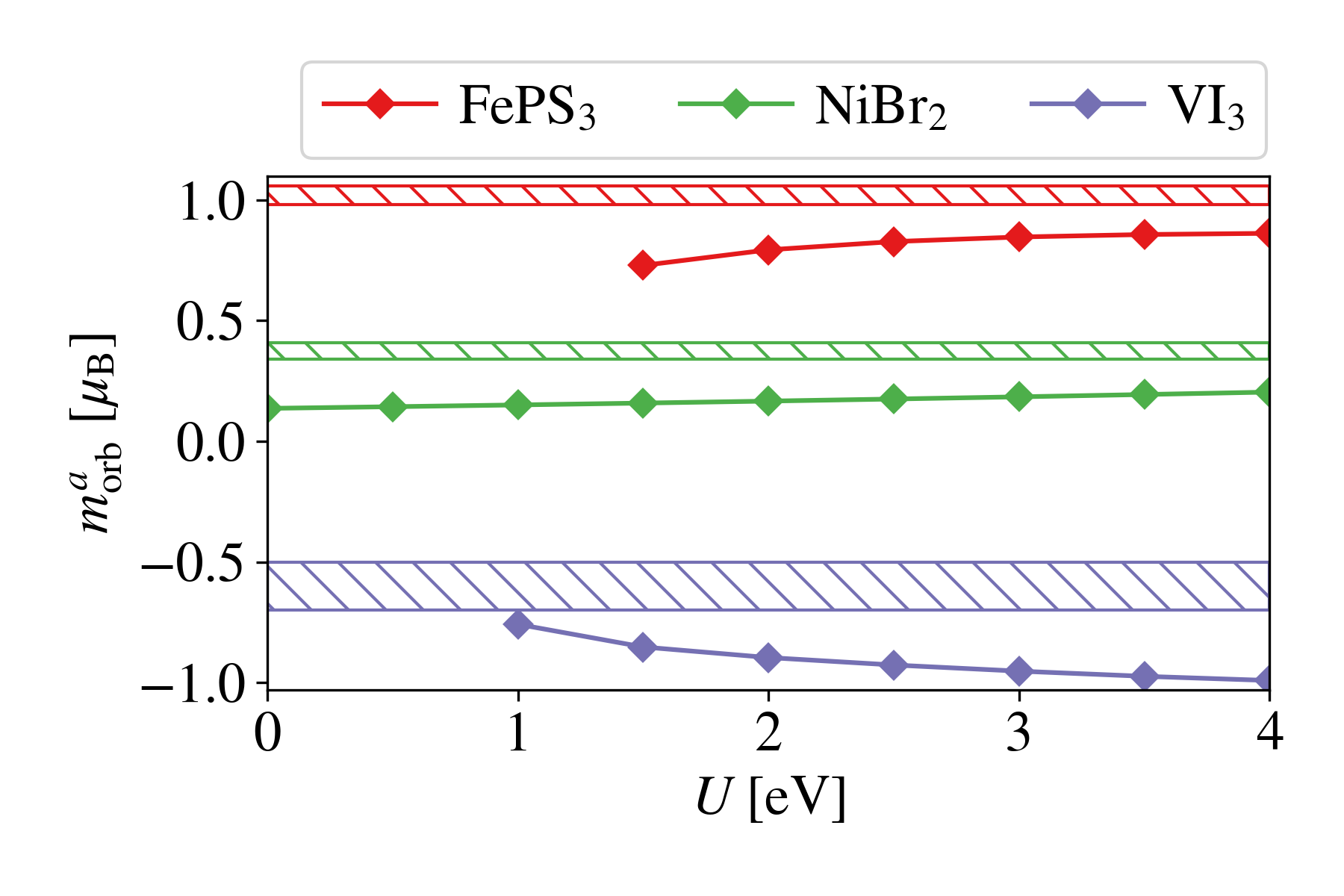}
    \caption{Calculated local orbital moments as a function of the Hubbard penalty $U$ for the crystals FePS$_3$, NiBr$_2$ and VI$_3$. Each point represents an SCF calculation and the hatched areas represent the range of the measured moments from \autoref{tab:vdw-compare}.}
    \label{fig:UvsOM}
\end{figure}

In order to abate these significant discrepancies, we turn to the Hubbard-like DFT+U correction as formulated by \citet{dudarev1998electron}, which adds a penalty functional with strength $U$ to the Kohn-Sham Hamiltonian. However, although DFT+U may open the gap in these materials the NSCF SOC orbital moments remain small. It should be noted though that NSCF SOC calculations assume that the spin-density is not changing upon inclusion of SOC. In the case of large orbital moments, this assumption is likely to be unjustified and the calculations have to be performed with self-consistent SOC. We note that due to the formal issues with functional derivatives in non-collinear DFT calculations \cite{komorovsky2019four}, we restrict ourselves to the local spin-density approximation (LSDA) functional  \cite{von1972local} in the locally collinear approximation \cite{kubler1988local}. The results for various values of $U$ are displayed in \autoref{fig:UvsOM} and the results for $U=3$ eV are presented in \autoref{tab:vdw-compare}. We see a dramatic improvement for FePS$_3$ and VI$_3$ and a minor improvement for hexagonal NiBr$_2$ when SOC is included self-consistently (in conjunction with Hubbard correction). Nevertheless, the orbital moments on the chromium compounds remain fairly small and aligned with the spin moments.

We emphasize that the improvement in FePS$_3$ and VI$_3$ cannot simply be attributed to either DFT+U or SCF SOC, but instead to the interplay between these two corrections. As displayed in \autoref{tab:vdw-compare}, collinear DFT+U calculations followed by NSCF SOC leads to small orbital moments. On the other hand, it is not possible to open a gap in either of the systems with DFT using self-consistent SOC without Hubbard corrections. For the case of FePS$_3$ the large orbital moments (requiring DFT+U and SCF SOC) was first noted by \citet{kim2021magnetic} whereas the large moment in VI$_3$ was predicted by \citet{Yang2020}.

The appearance of large orbital moments can be understood from simple crystal field splitting in conjunction with SOC. The case of NiBr$_2$, VI$_3$ and FePS$_3$ all exhibit octahedral field splitting where the $d$-band is separated into a $t_{2g}$ triplet and an $e_g$ doublet. Filling either of these does not yield any orbital angular momentum (in an atomic picture). Neither does partial filling of the $e_g$ band. In contrast partial filling of the $t_{2g}$ band can yield an angular momentum of $m_l=\pm1$ by forming linear combinations of the $d_{xz}$ and $d_{yz}$ orbitals. Without SOC, however, the bands are roughly degenerate and leads to a metallic state at partial filling. The Hubbard $U$ correction helps lift the degeneracy while SOC enforces a finite angular momentum in the occupied state. This is clearly the case for V atoms with $S=1$ (as in VI$_3$), which give rise to a splitting of the $t_{2g}$ band with two states below the Fermi level carrying a unit of angular momentum oppositely aligned with the spin. Similarly, for the $S=2$ state of Fe atoms in FePS$_3$, the majority $d$-bands are completely filled and the minority $t_{2g}$ band is occupied by one electron. With the inclusion of Hubbard corrections and self-consistent SOC, the $t_{2g}$ band splits up, leaving one occupied minority spin state with one unit of anti-aligned (aligned with majority spin) angular momentum. Finally, the Ni atoms in NiBr$_2$ have $S=2$ with completely filled majority bands as well as the minority $t_{2g}$ band and large orbital moments are not expected. The improvement of self-consistent SOC calculations with Hubbard corrections for NiBr$_2$ are thus likely related to the improved description with LDA+U rather than a {\it qualitatively} different electronic structure as in the case of VI$_3$ and FePS$_3$.

For materials with partially filled $t_{2g}$ bands, the results of the high-throughput calculations with NSCF SOC are not always expected to be reliable. In particular, if the onsite Coulomb interactions are strong enough to split the $t_{2g}$ band such that the monolayer becomes insulating, SCF SOC will tend to yield a large unquenched orbital momentum that cannot be obtained with NSCF SOC. A systematic treatment of such cases would require DFT+U calculations for the entire C2DB with SCF SOC, which is beyond the scope of this work. However, based on the crystal field arguments it should be straightforward to judge whether a prediction from NSCF SOC is reliable or if additional (more rigorous) calculations are required to verify the result. For other types of crystal field splitting similar arguments are applicable.
\begin{figure}[t]
    \centering
    \includegraphics[width=0.97\columnwidth]{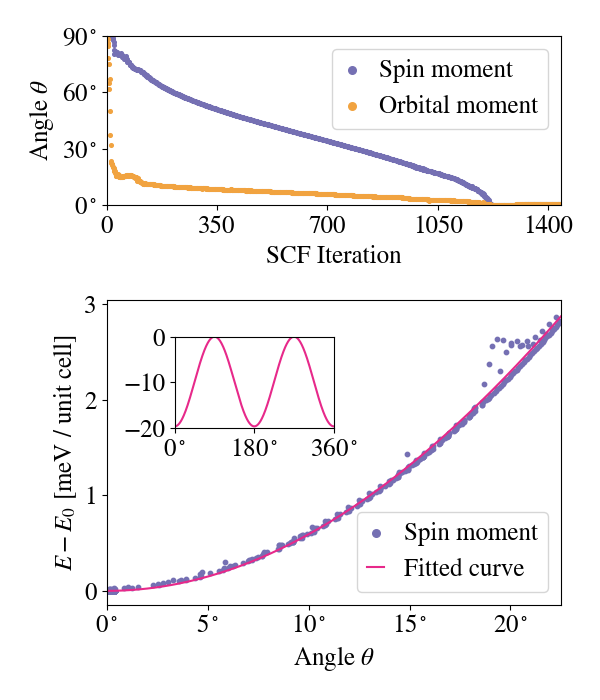}
    \caption{Top: The angle between localized spin and orbital moments and the out-of-plane axis for an Fe atom in the FePS$_3$ monolayer during the self-consistency cycle. Bottom: Scatter plot of the energy per unit cell of the last 500 SCF iterations and the angle between spin moments and the out-of-plane axis of the corresponding iterations. The data points have been fitted to a $\cos^2\brB{\theta}$ curve (full curve shown in inset) in order to retrieve $K$ from \eqref{eq:anisotropy}; the least-squares fit yields $K=9.8$ meV}
    \label{fig:FePS3_magani}
\end{figure}

Lastly, we wish to comment on the out-of-plane magnetic anisotropy of the FePS$_3$ monolayer, which has been shown experimentally to be between 11.2 and 12.0 meV per Fe atom \cite{lee2023giant}. The anisotropy may be modeled by a uniaxial single-ion anisotropy term $-AS_z^2$ in a Heisenberg model treatment. In a classical treatment this is expected to give rise to the energy
\begin{equation}
    E_\mathrm{anisotropy}=-2K\cos^2\brB{\theta},
    \label{eq:anisotropy}
\end{equation}
where $\theta$ is the angle between the direction of the spin moments and the out-of-plane axis and the factor of 2 takes into account the two Fe atoms in the unit cell. Since the Fe atoms have $S=2$, the out-of-plane magnetic anisotropy $K$ corresponds to an experimental single-ion anisotropy strength $A=K/S^2$ between 2.8 and 3.0 meV. Some previous attempts at calculating the single-ion anisotropy for this system yielded $A=0.101$ meV per Fe atom \cite{olsen2021magnetic}; a significant underestimation which is caused by the reliance of the magnetic force theorem (and NSCF SOC) during energy mapping analyses  \cite{kim2021magnetic}.

An accurate evaluation of the anisotropy in FePS$_3$ requires a ground state with the proper unquenched orbital angular momentum. This can be obtained from calculations including self-consistent SOC, but in such calculations the anisotropy energy can only be obtained using constrained DFT. We will refrain from doing such calculations here and simply resort to a rather crude approximation where an LSDA+U SCF SOC calculation is initialized with Fe spin moments perpendicular to the easy-axis. The evolution of orbital and spin angular momenta can then be tracked during the self-consistency cycle and recorded along with the energy. Since the electron density converges much faster than the direction of the spins, the relationship between energy and the direction of the spin moments for the last iterations of the SCF cycle can be fitted to the anisotropy curve in \eqref{eq:anisotropy} in order to determine $K$.

\autoref{fig:FePS3_magani} shows how the directions of the spin and orbital moments for an Fe atom change during the SCF calculation. The spin moments were initialized in-plane and their direction slowly changes until they are aligned with the easy-axis at $\theta=0$. The direction of the orbital moments, however, converges much faster and becomes aligned with the out-of-plane direction rather rapidly. The orbital magnetic moment is thus pinned to the $z$-axis rather than the spin, which gives rise to a significant magnetic anisotropy. We also show the energies of the last 500 iterations (measured from the total ground state energy $E_0$) as a function of the spin moment angle $\theta$. A least-squares fit between these data points and the anisotropy relation in \eqref{eq:anisotropy} yields $K=9.8$ meV per Fe atom such that $A=2.45$ meV, which is in much better agreement with experiments than the estimate from NSCF SOC. While these calculations represent a rather crude way of estimating the anisotropy without constraining the spin, our results are in good agreement with previous constrained DFT calculations, which also found an orbital moments being pinned to the lattice rather than the spin \cite{kim2021magnetic}.